%% file: main.tex
\documentclass[11pt]{scrartcl}  \usepackage[margin=2.0cm
]{geometry}

\usepackage[utf8]{inputenc}
\usepackage[T1]{fontenc}
\usepackage[utf8]{inputenc}
\usepackage{graphicx} % Required for inserting images
\usepackage{authblk}
\usepackage{caption}
\usepackage{gensymb}
\usepackage{subcaption}

\usepackage{blindtext}
\usepackage{subfiles} % Best loaded last in the preamble

\usepackage{siunitx}
\usepackage[backend=biber, style=numeric-comp,
, sorting=none, maxcitenames=99, maxbibnames=99]{biblatex}
\usepackage{booktabs}
\usepackage{multirow}
\usepackage{listings}
\usepackage{glossaries}
\usepackage{tabularx}
\usepackage{float}
\usepackage{color,soul}

\usepackage{hyperref}
\usepackage{xr-hyper}

\usepackage{seqsplit}

\usepackage{adjustbox}

\newcounter{sisec}
\title{The Interoperability Challenge in DFT Workflows Across Implementations}

\author[1,$\dagger$]{Simon K. Steensen}
\author[2,$\dagger$]{Tushar Singh Thakur}
\author[1]{Manuel Dillenz}
\author[3]{Johan M. Carlsson}
\author[4,*]{Celso R. C. R{\^e}go}
\author[5]{Eibar Flores}
\author[3]{Hamid Hajiyani}
\author[6]{Felix Hanke}
\author[1]{Juan María García Lastra}
\author[4]{Wolfgang Wenzel}
\author[2,7]{Nicola Marzari}
\author[1]{Tejs Vegge}
\author[7,*]{Giovanni Pizzi}
\author[1,*]{Ivano E. Castelli}

\affil[1]{Technical University of Denmark (DTU), Department for Energy Conversion and Storage, 2800 Kgs. Lyngby, Denmark}
\affil[2]{Theory and Simulation of Materials (THEOS) and National Centre for Computational Design and Discovery of Novel Materials (MARVEL), École Polytechnique Fédérale de Lausanne (EPFL), CH-1015 Lausanne, Switzerland}
\affil[3]{Dassault Systèmes Germany GmbH, Am Kabellager 11-13, D-51063 Cologne, Germany}
\affil[4]{Karlsruhe Institute of Technology (KIT), Institute of Nanotechnology, 76344 Eggenstein-Leopoldshafen, Germany}
\affil[5]{Department of Sustainable Energy Technology, SINTEF Industry, 7491 Trondheim, Norway}
\affil[6]{Dassault Systèmes UK Limited, 22 Science Park, Cambridge CB4 0FJ, United Kingdom}
\affil[7]{PSI Center for Scientific Computing, Theory and Data, Paul Scherrer Institute, 5232 Villigen PSI, Switzerland}
\affil[*]{Corresponding authors}
\date{}

\begin{document}
%\glsdisablehyper

\maketitle

\begin{center}
  \vspace{-2.5em} % adjust vertical spacing if you like
  \normalsize\textbf{$\dagger$These authors contributed equally to this work:}\\[0.3em]
  \normalsize\textbf{Simon K. Steensen and Tushar Singh Thakur}
  \vspace{1.0em}
\end{center}

\begin{abstract}
Interoperability and cross-validation remains a significant challenge in the computational materials discovery community. In this context, we introduce a common input/output standard designed for internal translation by various workflow managers (AiiDA, PerQueue, Pipeline Pilot, and SimStack) to produce results in a unified schema. This standard aims to enable engine-agnostic workflow execution across multiple density functional theory (DFT) codes, including CASTEP, GPAW, Quantum ESPRESSO, and VASP. As a demonstration, we have implemented a workflow to calculate the open-circuit voltage across several battery cathode materials using the proposed universal input/output schema. We analyze and resolve the challenges of reconciling energetics computed by different DFT engines and document the code-specific idiosyncrasies that make straightforward comparisons difficult.
Motivated by these challenges, we outline general design principles for robust automated DFT workflows.
This work represents a practical step towards more reproducible and interoperable workflows for high-throughput materials screening, while highlighting challenges of aligning electronic properties, especially for non-pristine structures.
\end{abstract}

\section{Introduction}

Materials discovery is a cornerstone of technological progress. Historically, this endeavor has predominantly been experimental, guided by chemical intuition, as illustrated by breakthroughs in the field of batteries~\cite{zheng2018review, BorahReview}, solar cells~\cite{NREL}, and heterogeneous catalysis~\cite{Zaera2022}. 
In recent years, there has been a paradigm shift toward computational approaches, powered by high-throughput atomistic simulations and modeling \cite{saal2013materials, jain2013commentary}. 
This transition has given rise to a new generation of self-driving laboratories (SDLs) and Materials Acceleration Platforms (MAPs)~\cite{aspuru2018materials, vegge2025, MAPoverview}, encompassing efforts that range from purely simulation-driven discovery pipelines \cite{kahle2020high} to fully autonomous laboratories \cite{szymanski2023autonomous}, as well as hybrid autonomous frameworks \cite{vogler2024autonomous}.
These platforms are designed to address the vast search space of potential materials, navigating it more efficiently through intelligent automation and feedback loops that combine physics-based simulations with data-driven modeling in a closed-loop fashion \cite{tabor2018accelerating, oganov2019structure, pollice2021data}.
A unifying requirement across a diverse set of MAPs is the development of workflows that are not only automated and reproducible \cite{peng2011reproducible, allison2016reproducibility}, but also interoperable across diverse computational tools and infrastructures \cite{govoni2021code, flores2020materials, Bosoni_2024, evans2025datatractor}. 
While automation and reproducibility have seen notable progress, achieving interoperability, i.e., the "I" in the FAIR (Findable, Accessible, Interoperable, and Reusable) guiding principles \cite{wilkinson2016fair}, remains one of the key outstanding challenges \cite{mons2020fair}, even with recent notable efforts in aligning AiiDA~\cite{pizzi2016aiida}, Jobflow~\cite{jobflow2024} and pyiron~\cite{pyiron} in Ref.~\cite{janssen2025}.
The fragmentation in electronic structure methods further exacerbates this difficulty \cite{harrison2012electronic, helgaker2013molecular}, as well as the lack of unified standards for general data sharing practices \cite{tedersoo2021data}. 

Many of the automated workflows developed within the materials science community rely on Kohn-Sham density functional theory (DFT) \cite{hohenberg1964inhomogeneous, kohn1965self} as the underlying simulation method \cite{jain2016computational, marzari2021electronic}, and implemented in software such as CASTEP \cite{clark2005first}, GPAW \cite{mortensen2005real, mortensen2024gpaw}, Quantum ESPRESSO \cite{giannozzi2009quantum, giannozzi2017advanced}, VASP \cite{kresse1996efficient, PhysRevB.59.1758}, and many more.
However, the current landscape of these simulation engines is highly heterogeneous, characterized by distinct input/output formats and inconsistent parameter conventions.
This lack of standardization shifts the burden of interoperability onto the workflow managers, often necessitating substantial pre- and post-processing logic tailored to individual simulation engines. 
Ensuring reproducibility in DFT calculations is therefore not a trivial task, and has required considerable collaborative efforts across the electronic structure community \cite{lejaeghere2016reproducibility,Bosoni_2024}. 
Despite the overall positive outlook on reproducibility by Lejaeghere \textit{et al.}~\cite{lejaeghere2016reproducibility}, they (and similarly Bosoni \textit{et al.}~\cite{Bosoni_2024} in their expanded work) mainly focused on total energies and equation of states of pristine materials \cite{lejaeghere2016reproducibility, lejaeghere2014error}. 
Investigating more complicated interactions, such as electron-phonon coupling and zero-point motion renormalization, can be highly non-trivial \cite{ponce2014verification}, and simply using an identical set of input parameters across different DFT engines often does not yield identical results \cite{carbogno2022numerical}. 
At a minimum, a set of well-defined guiding principles, which are code-aware but workflow-transparent, is required when comparing material properties derived from simulations run on different DFT engines \cite{Bosoni_2024}, and then building upon them for further alignment.

As noted previously, several workflow management systems have emerged in recent years, providing essential infrastructure for managing and automating electronic structure calculations \cite{shahzad2024accelerating}. 
Prominent examples include the workflow engines AiiDA~\cite{pizzi2016aiida, huber2020aiida, uhrin2021workflows}, Jobflow~\cite{jobflow2024}, Taskblaster~\cite{taskblaster2025}, pyiron~\cite{pyiron}, MyQueue~\cite{mortensen2020myqueue}, PerQueue~\cite{sjolin2024perqueue}, the proprietary BIOVIA Pipeline Pilot~\cite{BPP}, and SimStack~\cite{Rego_2022}. 
While these systems have advanced the state of automation and reproducibility, they differ significantly in how they interface with various DFT engines and in handling the flexibility and idiosyncrasies inherent to each code. 
Effectively, addressing these interoperability challenges is crucial for developing scalable, engine-agnostic workflows that can adapt to rapidly evolving research needs without compromising scientific rigor.

% would be good to have some sketch showing workflow managers interfacing with DFT engines

Addressing interoperability challenges within the DFT-enabled materials discovery ecosystem requires progress on two key fronts.
First, the definition and adoption of a machine- and human-readable universal input/output standard that can be translated internally by workflow managers into engine-specific formats. 
This standard must be rich enough to represent the full complexity of electronic structure calculations and modular enough to support transparent translation to code-specific inputs. 
Second, there must be a concerted effort to ensure that nominally identical inputs yield quantitatively comparable results across engines, with well-established tolerance criteria and correction strategies in place.

In this study, we focus on solving the first challenge of interoperability while building on existing efforts that have thoroughly addressed the second concerning DFT implementations~\cite{lejaeghere2016reproducibility, Bosoni_2024}. 
Specifically, we build on the workflow proposed by B{\"o}lle \textit{et al.}~\cite{Bolle_2020} to compute open circuit voltages (OCV) across a range of battery materials (typically a potential cathode material) and states of charge (SOC), and we put a particular emphasis on the analysis of the alignment of OCV values across different DFT engines. 
We implemented this workflow across several workflow managers, utilizing a unified input/output standard designed and developed by us to enable cross-code interoperability.
The details of this implementation, along with the design of the standard schema, are presented in Section \ref{method}.
A detailed discussion of the OCV results, including the challenges in ensuring consistent voltage predictions, is given in Section \ref{sec:result}.
Drawing from these challenges, we outline general design guidelines for automated workflow design, with an emphasis on the benefit of cross-code validation.
Finally, in Section \ref{conclusion}, we summarize our findings, followed by an outlook on addressing the remaining challenges to developing interoperable workflows and the broader implications for reproducibility in high-throughput materials discovery. 

\section{Methods} \label{method}

\subsection{Common input and output standards} \label{io-schema}

A fundamental step towards achieving workflow interoperability lies in the definition of input and output (I/O) formats, which enable seamless data exchange across different implementations. 
To support modularity and scalability in complex scientific workflows, we have developed a unified I/O schema, inspired by the work in Ref.~\cite{Huber2021}, based on the data serialization format JSON and, by definition, its superset YAML.
This standardized format ensures consistency by explicitly and exhaustively defining the input each engine receives and the output it returns. 
The definitions are formalized as JSON schemas, which facilitate direct integration with APIs, simplifying possible integration within the broader context of MAPs.
Throughout this work, practical insights emerged regarding which parameters must be explicitly defined in the I/O formats and which need to be set internally by the workflow managers, particularly in response to the performance characteristics and conventions of the underlying DFT engines.

\paragraph{Input format} The input schema is organized into mandatory and optional fields to balance standardization with flexibility. 
Among the mandatory fields are: 1) the atomic structure, expressed in compliance with the OPTIMADE specification~\cite{andersen2021optimade,Evans2024}, 2) a protocol keyword (currently supporting ``fast'', ``moderate'', and ``precise''), which governs the internal parameterization of the DFT engine based on the desired trade-off between precision and computational cost, and 3) the execution context, specifying the computational environment or location of the high-performance computing (HPC) resource.

The optional inputs enable fine-grained control over simulation parameters that are otherwise internally managed by selecting a protocol. 
This design ensures that all relevant DFT parameters remain transparent and user-accessible, supporting a high degree of customizability when needed. 
A non-exhaustive list of examples includes the k-point mesh density, spin polarization settings, constraints on volume change between charged and discharged states, definition of charge carriers in the electrode material, and definition of supercell size construction.

Finally, a dedicated optional meta-configuration block is included to capture engine-specific or infrastructure-related details. 
These inputs accommodate cases such as selecting a particular quantum code if multiple options exist, defining custom computational resource allocations, toggling engine-specific features that may not be possible on other engines, and ensuring compatibility with both general-purpose and specialized simulation workflows.

\paragraph{Output format} In a similar fashion, the output schema is structured into mandatory and optional fields to accommodate both standardized reporting and engine-specific details.
The mandatory outputs include: 1) a structured dictionary containing the OCV values computed at the high and low SOC, along with the average OCV, and 2) all relaxed structures generated during the workflow, represented in accordance with the aforementioned OPTIMADE standards. 
These core outputs provide a consistent and minimal set of results required for downstream analysis and comparison across workflows and engines.

The optional outputs expose additional simulation details that may be useful for diagnostic purposes or a more detailed investigation.
These include total energies, atomic forces, stress tensors, and magnetic moments that may not always be required depending on the specific use case.
By structuring these optional fields in a modular and extensible manner, the output schema remains flexible while preserving a high level of semantic clarity and interoperability.

A complete definition of mandatory and optional fields for both input and output schema is provided in the Supplementary Information~\ref{SI:sec:IO}.

\subsection{Workflow} \label{workflow}

\begin{figure}
    \centering
    \includegraphics[width=\linewidth]{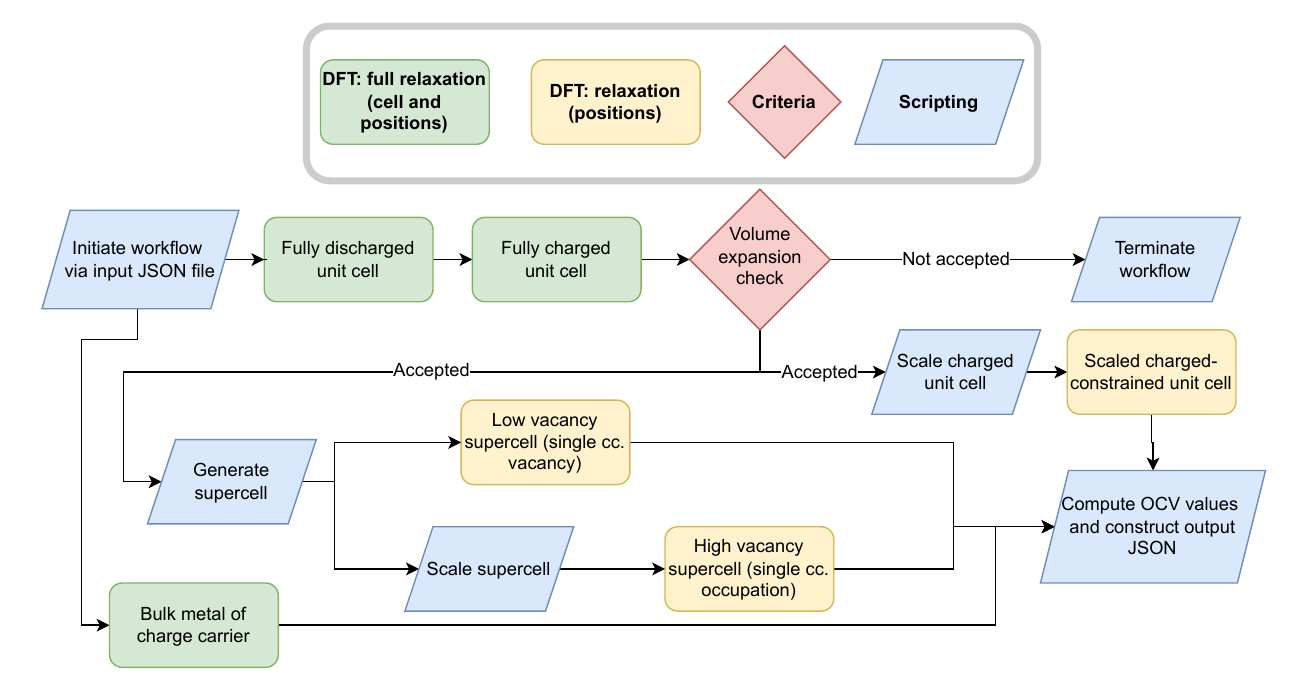}
    \caption{The workflow begins by reading the input JSON and ends by constructing the output JSON. DFT stages comprise a full cell\mbox{+}position relaxation of the discharged unit cell followed by a relaxation of the charged cell. A single termination criterion bounds the relative volume expansion between charged and discharged states to a user-specified limit (default \(5\%\)); if the check fails, the charged-constrained cell is produced through scaling. Upon acceptance, the workflow generates and scales supercells, evaluates low- and high-vacancy configurations (single \emph{cc.} vacancy/occupation), and outputs OCV values along with metadata. Here, \emph{cc.} abbreviates charge carrier.}
    \label{fig:workflow}
\end{figure}

The intercalation workflow illustrated in Figure~\ref{fig:workflow} initializes from a fully intercalated unit cell of a known or hypothetical cathode material.
The initial step involves relaxing the discharged state.
%, thereby establishing a mechanically stable reference configuration. 
A corresponding charged state is then generated by removing all charge carriers (henceforth referred to as cations) and relaxing the resulting structure. This enables a direct calculation of the volume change descriptor, a key metric for assessing structural reversibility during battery cycling, 
\begin{equation} \label{eq1}
\Delta V = \frac{V_{charged} - V_{discharged}}{V_{discharged}},
\end{equation}
where $V_{charged}$ and $V_{discharged}$ are the volumes of the charged and discharged unit cells. A termination criterion in the workflow, based on a volume change limit, is provided in the designed input structure, illustrated as the red square in Figure \ref{fig:workflow}. 

The average OCV value is then computed by
\begin{equation} \label{eq2}
V_{average} = \frac{(E_{charged} + N_{ion} \cdot E_{bulk-cation} / N_{bulk-cation}) - E_{discharged}}{N_{ion}\cdot z}
\end{equation}

where $E_{charged}$ and $E_{discharged}$ are the total energies (in eV) of charged and discharged structures respectively, $E_{bulk-cation}$ is the total energy (in eV) of the working cation in its most stable bulk metallic structure with $N_{bulk-cation}$ atoms, $N_{ion}$ is the difference between the number of working cations present in the fully discharged and fully charged structures, and $z$ is the oxidation state of the working cation when intercalated in the cathode. We ensure that the structure is the most stable by verifying that the energy of the metallic bulk system lies on the convex hull of formation energies, which demonstrates thermodynamic stability.

The workflow continues to create the charged-constrained structure. This is constructed by first retrieving the lattice lengths along the $x$, $y$, and $z$ directions for the discharged and charged unit cells, and then calculating the fractional changes in each direction. The discharged structure is then copied and rescaled according to these directional changes, and finally, the charged carriers are removed to form the charge-constrained structure. This structure is then relaxed with the lattice parameters fixed.
Note that this implementation extends the original approach by increasing physical accuracy. Rather than uniformly scaling all directions based on the total volume change as done in the original workflow, each lattice direction is scaled independently.

% These three relaxed structures: discharged, charged, and charged-constrained, form a core structural triplet that underpins the remainder of the workflow and is constructed automatically, independent of the DFT engine employed.

%ensuring a minimum defect spacing of 8 \AA between opposite faces to minimise spurious interactions from periodic images. 

Simultaneously, as the charged-constrained structure is generated, the workflow generates a supercell from the discharged configuration.
In this study, the five investigated materials were assigned fixed scaling matrices for 1:1 comparability. To preserve computational tractability, the workflow caps the generated supercell at 150 atoms. This limit generally suits the cathode materials relevant for the workflow, avoiding unnecessary computational cost while still providing a typical defect–defect spacing of at least 8~\AA (or close to) to mitigate spurious interactions from periodic images.
A symmetry analysis is then performed to identify inequivalent cation sites. Each of these sites is selectively emptied to produce dilute-vacancy configurations, representing the low-SOC configuration. 
Conversely, the high SOC limit is modeled by retaining a single cation in the charged lattice and applying the same anisotropic rescaling to the charged-constrained case.
%This design enables the simultaneous assessment of both structural stability and thermodynamic driving force in a single, fully automated pass. 
With these structure relaxed, the low SOC
\begin{equation} \label{eq3}
V_{low-SOC} = \frac{(E_{low-SOC} + N_{ion} \cdot E_{bulk-cation}/ N_{bulk-cation}) - \alpha \cdot  E_{discharged}}{N_{ion}\cdot z},
\end{equation}
and high SOC 
\begin{equation} \label{eq4}
V_{high-SOC} = \frac{(\alpha \cdot E_{charged-constrained} + N_{ion} \cdot E_{bulk-cation}/ N_{bulk-cation}) - E_{high-SOC}}{N_{ion}\cdot z}
\end{equation}
are computed, with $E_{low-SOC}$, $E_{charged-constrained}$, and $E_{high-SOC}$ being the energy of the low SOC supercell, constrained charged cell, and high SOC supercell, respectively. $N_{ion}$ is the ion discrepancy between the two states compared with respect to the scaling parameter $\alpha$ defined as $N_{atoms-in-supercell} / N_{atoms-in-unitcell}$.

%where $E_{charged-constrained}$ and $E_{low-SOC}$ are the total energies (in eV) of the charged-constrained unit cell and high SOC supercell, $N_{ion}$ is the ion discrepancy between charged-constrained and high SOC structures, which in this case is simply one.

The entire workflow is expressed in a declarative manner rather than as a fixed imperative sequence. 
Once an input structure is provided, a workflow manager can execute the same directed acyclic graph of tasks, translating the universal scientific intent into engine-specific instructions while automatically capturing provenance information in compliance with FAIR data principles. 
%By decoupling the scientific logic from infrastructural idiosyncrasies, this approach ensures engine agnostic, cross platform automation, which is essential for enabling interoperable, and scalable screening of intercalation electrode materials.

\subsection{Implementations}

The workflow outlined in Section~\ref{workflow} has been implemented in four representative and widely adopted workflow managers: AiiDA, PerQueue, Pipeline Pilot, and SimStack, each coupled to an appropriate plane-wave DFT engine and queued to its target HPC environment. Across these implementations, the core logic of Figure~\ref{fig:workflow} is preserved (JSON I/O, sequential and parallel DFT relaxations, supercell generation, computing OCV values, etc.), while engine and site-specific adapters handle submission, provenance, and data marshaling. Below, we summarize the distinctive capabilities of each framework and how they enable the scalable and interoperable execution of the OCV protocol in heterogeneous computational ecosystems.

\paragraph{AiiDA integration utilizing Quantum ESPRESSO:}AiiDA (Automated interactive infrastructure and database for computational science) is a Python-based infrastructure and workflow manager for reproducible high-throughput simulations \cite{pizzi2016aiida, huber2020aiida, uhrin2021workflows}. A key strength is its comprehensive provenance tracking: it records the full lineage of every data object and computational step, including all inputs, metadata, and dependencies. This provenance is stored as a directed acyclic graph in a relational database, with nodes representing data objects or computational processes \cite{huber2022automated}.
% Fig. \ref{} illustrates this capability in an acyclic graph, taken from this work, that illustrates the entire OCV workflow for one structure. 

The OCV workflow is implemented within an open-source Python package as an AiiDA plugin, called \texttt{aiida-open\_circuit\_voltage} \cite{aiida-ocv}, which builds upon the established \texttt{aiida-quantumespresso} plugin \cite{aiida-qe}. 
Beside adhering to the I/O structure the \texttt{aiida-open\_circuit\_voltage} plugin adheres to AiiDA's native format of storing inputs and outputs as nodes in a local SQL database. 
Computational jobs are automatically submitted to the HPC environments, with the plugin managing resource tracking, job submission, and most importantly, failure recovery in a fully automated manner. 
Beyond the native error handling features provided by the \texttt{aiida-quantumespresso} plugin, we have implemented a set of custom error handlers tailored specifically to the OCV workflow. 
These additions ensure robust execution by autonomously detecting and recovering from common failures, minimizing human intervention and supporting high-throughput screening.

\paragraph{PerQueue integration utilizing VASP:} A Python-based workflow has been implemented using the dynamic workflow manager PerQueue and the Vienna Ab Initio Simulation Package (VASP, v6.3.2). PerQueue is designed for constructing complex workflows and supports ad hoc job submission, where workflow actions are dynamically adjusted based on intermediate results.
Internally, PerQueue generates a graph representation of the specified workflow using networkx \cite{networkx}. This design enables switching executable paths, iterating within closed loops, and expanding the width of the workflow on the fly (e.g., launching $N$ jobs, where $N$ is later determined by intermediate results), with all behavior governed by the workflow’s traversal of the graph. The workflow manages and submits jobs across an HPC system, with the possibility to run smaller jobs directly on the login node upon workflow initialization. Dynamic capabilities of workflow managers, such as PerQueue, enable robust workflow implementation and have been argued to facilitate the integration of self-driving labs with higher complexity~\cite{Steensen2025Perspective}.
The implementation utilizes ASE \cite{ase-paper, ISI:000175131400009} both to define the required structures and to launch VASP (v6.3.2) for structural relaxations.
Each calculation produces its own JSON file in the OPTIMADE structure format, which facilitates generating a final aggregated JSON output and simplifies debugging and tracking of individual calculations.

\paragraph{PerQueue integration utilizing GPAW:} Using a setup analogous to the PerQueue/VASP design, ASE enabled a straightforward extension of the workflow to GPAW (v24.1.0) using the same skeleton. The necessary adaptations involved implementing the relaxation procedure (internal ASE implementation used) and adjusting file names and formats in the workflow. Only DFT-specific components have been modified for GPAW.

\paragraph{Pipeline Pilot integration utilizing CASTEP:} The workflow using the CASTEP code~\cite{clark2005first} is implemented in BIOVIA Pipeline Pilot (BPP), which is a general modular environment for data science and automation~\cite{BPP}. Each component executes elemental tasks like reading or parsing incoming data, executing computational tasks, analyzing data and generating reports. The components can be connected via data pipelines to generate complex workflows called protocols.  
The CASTEP code~\cite{clark2005first} is a computational module in BIOVIA Materials Studio package~\cite{MS}. 
The functionality in the Materials Studio has been modularized into components in the Materials Studio Collection for Pipeline Pilot, such that computational workflows can be built easily. 
The BPP protocol implementing the OCV workflow starts by reading the JSON-input file and generating unit cells of the target material and the charge carrier in Materials Studio format using Materials Script, which is a perl-based scripting language.
The unit cell of the charge carrier, the charged and discharged unit cells are optimized using the CASTEP DFT component in parallel pipelines. The optimized cells are used to generate the scaled and defect supercell in further parallel pipelines according to the OCV workflow.
The results of the six pipelines are finally gathered together to calculate the OCV values using the intrinsic Pilot Scripting language. 
 
\paragraph{SimStack integration utilizing VASP:}
SimStack \cite{Rego_2022, Schaarschmidt_2021, Bekemeier_2025} is a lean, SSH-only client-server framework that enables researchers to orchestrate workflows locally while delegating execution to remote HPC resources. On “run”, the client bundles the XML description, helper scripts, and user data into a self-contained package, which is streamed to the target machine. The server then spawns one job per Workflow-Active-Node (WaNo), routes inter-node data, and monitors execution without requiring admin privileges or scheduler hacks. Each WaNo is defined by a compact XML template that unifies the run command, I/O contract, and GUI controls, enabling complete protocols to be shared or archived by distributing the XML along with companion Python scripts, ensuring no hidden state. In practice, this means that a protocol assembled on a laptop can be replayed byte-for-byte on a supercomputer. The same mechanism also simplifies collaboration, since exchanging a WaNo preserves both parameters and interface semantics. This yields end-to-end provenance by construction, making outputs immediately FAIR and reproducible across clusters with a compatible VASP binary and a standard Python stack~\cite{Soleymanibrojeni_2024, Mostaghimi_2022, Dalmedico_cover_2024, Dalmedico_2024, DaSilva_2025, Mieller_2024, deAraujo_2024, Pecinatto_2023}.

The \texttt{DFT-VASP} WaNo \cite{DFT-VASP, Rego_2025} at \url{github.com/KIT-Workflows/DFT-VASP} illustrates this: its template exposes physically meaningful DFT knobs (cell vectors, smearing, \texttt{ENCUT}). At the same time, boilerplate file handling is delegated to ASE, pymatgen, NumPy, and PyYAML. A four-tab interface streamlines setup (INCAR with inline tooltips; KPOINTS with length-based or Monkhorst meshes; Files--Run selecting \texttt{vasp\_std}, \texttt{vasp\_gam}, or \texttt{vasp\_ncl} and auto-generating \texttt{POTCAR} once \texttt{POSCAR} is provided). At runtime, the node fabricates control files, launches VASP, and condenses outputs into \textit{vasp\_results.yml}, linking energies, forces, and magnetic moments unambiguously to inputs. The same drag-and-drop paradigm scales from single calculations to high-throughput campaigns by swapping WaNos, chaining pre/post-processing, and distributing parameter sweeps, without touching a shell script~\cite{Bekemeier_2025}.

\subsection{Semantic Description}
Given the variety of datasets, formats, software, and processes implemented, we additionally provide a metadata description of the workflow using modern semantic web technologies~\cite{SemanticWeb, clark2022toward}. Concretely, we systematically describe the workflow (inputs, methods, and results) as a graph of objects and their attributes, serialized into JSON-LD format. The attributes, i.e., keys, are not created arbitrarily; instead, they are reused from controlled vocabularies and ontologies. Concepts from scientific domains are inherited from the EMMO~\cite{EMMO}, BattINFO~\cite{BattINFO, clark2025semantic, clark2022toward} (developed under the BIG-MAP consortium~\cite{BIG-MAP}), and OSMO~\cite{OSMO} ontologies, while resource descriptions are inherited from widely used web vocabularies such as Schema.org~\cite{guha2016schema}. A machine-readable description of the workflow improves its data FAIRness; i.e., the JSON-LD description renders the workflow indexable and thus Findable in knowledge bases, while using domain-specific vocabularies improves its Interoperability across research domains. A more detailed description of the semantic description can be found in the Supplementary Information section~\ref{SI:sec:sematics}.

\section{Results and discussion} \label{sec:result}
We present the findings on the alignment and interoperability of OCV calculations across the different DFT implementations, with a particular focus on the practical challenges that arise in automated high-throughput materials discovery workflows. The materials under investigation here are representative battery electrode materials: Li$_2$Mn$_3$NiO$_8$ (Materials Project ID mp-771112, 56 atom unit cell), LiFePO$_4$ (mp-19017, 28 atom unit cell), LiTiS$_2$ (mp-9615, 4 atom unit cell), LiCoO$_2$ (mp-849273, 64 atom unit cell%, note not layered oxide
) and MgMo$_3$S$_4$ (mp-677217, 16 atom unit cell). These materials span a broad range of transition metal chemistries to test the workflow implementations, including elements known to challenge pseudopotential representation~\cite{Bosoni_2024}.

We begin by discussing the primary sources of systematic deviation and the remaining challenges under standardized conditions. Following this, we summarize the degree of agreement achieved between different computational codes shown in Figure~\ref{fig:alignment} and explore the broader implications for modular computational workflows in material discovery. The scaling matrices used for supercell construction are detailed in SI Section~\ref{SI:scaling}.

%%%%%%%%%%%%%%%%%%%%%%%%%%% Input challenges %%%%%%%%%%%%%%%%%%%%%%%%%%%%%%%%%%%%
\subsection{DFT alignment challenges}
All implementations started by agreeing on key computational parameters: the PBEsol exchange-correlation functional, non-spin-polarization, a k-point density of 0.15~\AA$^{-1}$, and Fermi-Dirac smearing of 0.2~eV. However, other code-specific settings, such as plane-wave cutoffs, convergence criteria, symmetry operations, optimization algorithm and pseudopotentials, remained at the discretion of each developer, following best practices. The scaling matrices used for supercell construction are detailed in SI Section~\ref{SI:scaling}.
Despite these efforts, substantial discrepancies for several of the OCV results persisted, necessitating further systematic and thorough analysis of parameter choices.

%Our initial tests using the moderate settings showed large variations and it was required to use precise settings to bring the results of the different implementation towards better agreement. This indicates that DFT codes may approach minimizatin through different optimization paths ways and only fully converged results are comparable across codes. 

% To diagnose these discrepancies, we conducted an in-depth comparison between the results obtained from VASP and Quantum ESPRESSO. LiCoO$_2$, LiFePO$_4$, and MgMo$_3$S$_4$ were studied first, with LiFePO$_4$ exhibiting the largest misalignment. This prompted an expansion of the dataset to a broader range of transition-metal chemistries by including Li$_2$Mn$_3$NiO$_8$ and LiTiS$_2$ to determine whether the deviations were specific to Fe, an element known to challenge pseudopotential methods across different DFT codes~\cite{Bosoni_2024}. This expansion revealed further misalignment issues for the introduced Li$_2$Mn$_3$NiO$_8$, indicating that LiFePO$_4$ was not the only challenging case, whereas the simpler LiTiS$_2$ remained well aligned upon integration.

%%% Smearing challenges
\subsubsection{Smearing effects on vacancy structures}\label{sec:AlignmentChallenges}
Cases of severe misalignment of Li$_2$Mn$_3$NiO$_8$ led us to conduct further systematic tests, focusing on the influence of smearing, k-point sampling and plane-wave cutoffs.
While k-point and cutoff convergence were already satisfactory, one of the most critical and sensitive parameters proved to be the smearing of electronic states at the Fermi level. It became apparent that energetically converged parameters on pristine unit cells do not guarantee converged parameters for non-pristine structures, such as the low- and high-SOC supercells. This is a general approach for conducting large-scale screening studies, as convergence testing on every single structure throughout the workflow is computationally expensive, making it important to keep in mind.

To systematically assess smearing effects, we ran the entire workflow for Li$_2$Mn$_3$NiO$_8$ using different smearing types and widths. Gaussian smearing widths were set to 0.05, 0.075, 0.10, 0.125, and 0.15 eV, and corresponding Fermi-Dirac smearing widths of 0.019, 0.029, 0.039, 0.049, and 0.058~eV were used.
The Gaussian smearing values were picked initially and the Fermi-Dirac values were calculated based on the scaling factor $({\sqrt{2/3}\pi)}^{-1}$ to make the results comparable to the Gaussian smearing width values as suggested in~\cite{Santos_2023}. Initially, both materials were tested at the lowest and highest smearing widths to determine if there were significant differences.
%For LiFePO$_4$, unlike Li$_2$Mn$_3$NiO$_8$, no additional alignment effects were observed within the tested range (see SI Figure~\ref{fig:LFP_smearing}), and therefore the intermediate widths were not investigated further.

\begin{figure}
    \centering
    \includegraphics[width=0.95\linewidth]{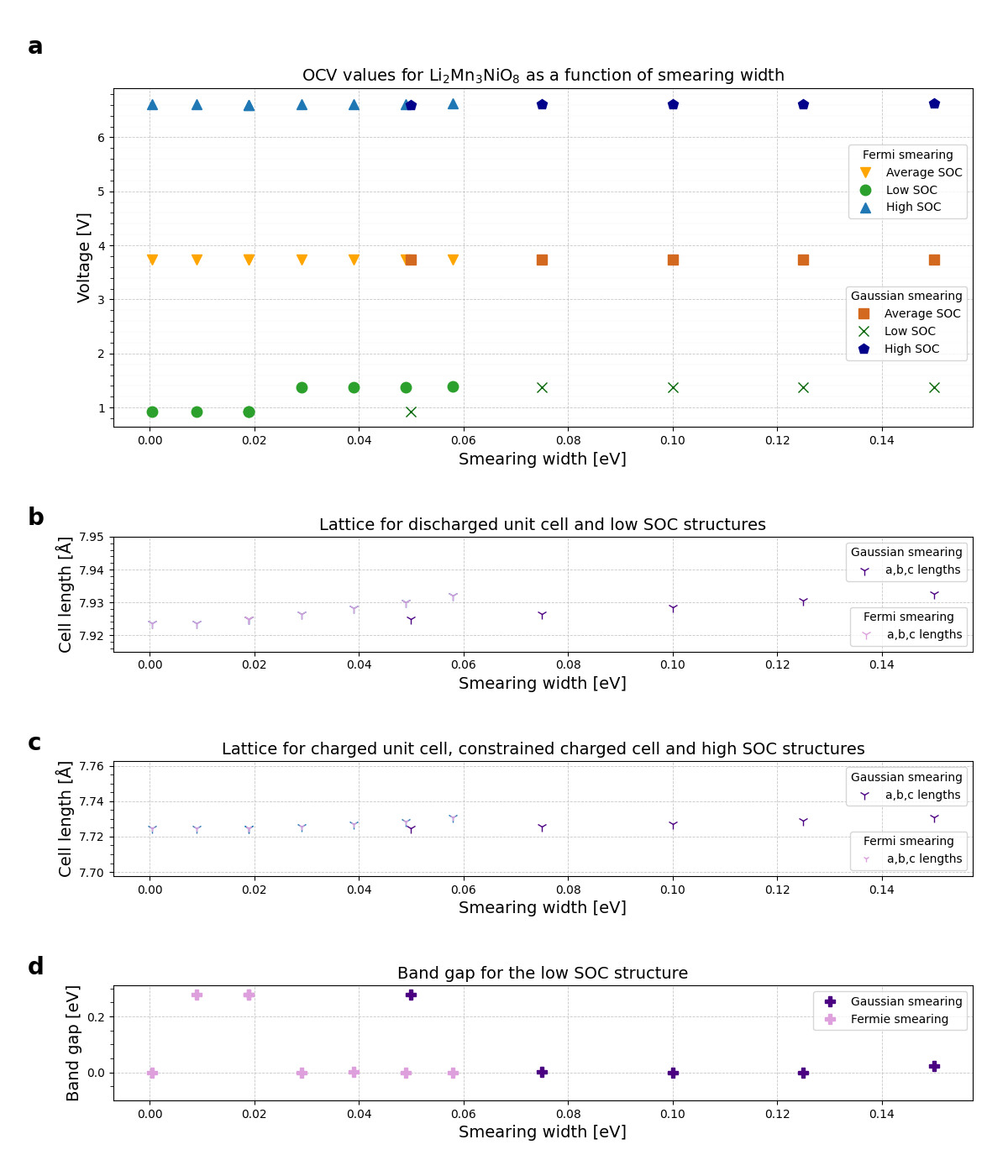}
    \caption{a) The plot illustrates the impact of smearing width (Gaussian and Fermi) on the OCV values for Li$_2$Mn$_3$NiO$_8$ using the PerQueue/VASP implementation. The average and high SOC are converged even at relatively high smearing. The low SOC exhibits a jump at low smearing. Note that two additional tests with Fermi–Dirac smearing widths of [0.005, 0.09] eV were conducted to ensure a convergence plateau is reached for the lattice parameters of the discharged unit cell. Plots b) and c) below show the smearing dependence of the a, b, and c lattice parameters for the simple cubic lattice. In the case of Li$_2$Mn$_3$NiO$_8$, no scaling is performed to construct the supercell, since it is deemed sufficiently large, meaning the discharged unit cell and low SOC structure will have identical lattice parameters, which is seen in the second plot. Furthermore, the cubic nature of both the discharged and charged unit cell yields identical cells for the charged and charged-constrained system, which is also identical to the high SOC structure. The lattice parameters of all three of these structures are therefore represented together in plots b) and c). d) show the bandgap extracted in the low SOC structure at the different smearing values and types. Here, an abrupt change in the opening of a bandgap is seen corresponding to the change in the low OCV value. Though with the band gap closing again at very low Fermi-Dirac smearing.}
    \label{fig:smearing_dependence_LNMO}
\end{figure}

The smearing study using the PerQueue/VASP implementation for Li$_2$Mn$_3$NiO$_8$ is shown in Figure~\ref{fig:smearing_dependence_LNMO}. Although OCVs at high and intermediate SOCs remain relatively stable, the low SOC voltage exhibits complex non-monotonic behavior that exemplifies the challenges of parameter convergence in automated workflows. While the average OCV and the OCV at high SOC remain relatively stable across smearing widths, the OCV at low SOC exhibits an abrupt drop between Fermi-Dirac smearing widths of $0.029$ eV and $0.019$ eV, despite the corresponding change in lattice parameter being only $0.0014$ Å.
% Additional calculations using a higher k-point density of 0.10~\AA$^{-1}$ yielded consistent results.
This discontinuous behavior originates from a smearing-induced artifact in the electronic structure rather than an actual metal-insulator transition. As seen in Figure~\ref{fig:smearing_dependence_LNMO}d, two near-degenerate states exist at the Fermi level whose fractional occupations critically depend on the smearing width. With minimal smearing ($0.0005$ eV), the sharp Fermi cut-off ensures only the lower state is occupied, yielding a negligible bandgap consistent with metallic behavior. At intermediate smearing values ($0.009-0.019$ eV), both states acquire fractional occupations, creating an apparent $0.27$ eV gap to the next manifold of states, a gap that exists only due to the partial occupation scheme. Further increasing the smearing causes even higher-energy states to be fractionally occupied, closing this artificial gap. Such behavior, characterized by abrupt rather than gradual changes, complicates workflow automation and convergence diagnostics. In principle, one could envision a procedure where each structure is tested with multiple smearing widths and convergence is automatically checked. However, as Figure~\ref{fig:smearing_dependence_LNMO} shows, even such procedures may fail to detect behaviors that do not converge towards the lower value in the intermediate values, as seen for the low SOC case. As seen in the Figure \ref{fig:PDOS_LNMO} in the SI, this does in fact correspond to an apparent change in the projected density of states of the two different energy levels, though relaxed with tight identical parameters. Note that spin-polarized calculations based on an exchange-correlation functional better suited for the description of strongly correlated \textit{d} electrons would likely lead to an opening of the bandgap and elevate this issue in this specific case.

%In the case of LiFePO$_4$, the VASP computed low SOC OCV value remained misaligned of about $\approx0.4$~V to the other DFT code after the smearing study, as seen in Figure~\ref{fig:alignment}. To make sense of this, the PDOS for the low SOC structure for Quantum ESPRESSO and VASP were generated as seen in the SI Figure~\ref{fig:PDOS_LFP}. The PDOS plots reveal no notable differences in the electronic structure, with both showing a similar band gap around the Fermi level and general structure.

%%% Results
\subsubsection{Influence of optimization procedure}
Beyond smearing, the challenge of identifying global instead of local minima through relaxation procedures can strongly influence the resulting structures and voltages which was greatly exemplified by further investigation into Li$_2$Mn$_3$NiO$_8$. For VASP and CASTEP, the initial discharged and charged Li$_2$Mn$_3$NiO$_8$ unit cells remained cubic (thereby all subsequent structures in the workflow), as the initial the Materials Project input structure. The corresponding voltages as seen in the summary Figure~\ref{fig:alignment}) are extreme at both low and high SOC, being around 1 V and 6.4 V, respectively. In contrast, in GPAW and Quantum ESPRESSO the cubic symmetry is broken yielding a similar average OCV, yet differences of several volts at low and high SOC.

To investigate this, we reran the entire VASP workflow using the GPAW relaxed discharged unit cell as the given structure in the input file. Note that while various procedures break the symmetry and identify global minima exist\cite{Mosquera-Lois2022}, these are not commonly implemented in the context of automated workflows.
Here, the full relaxations of the charged and discharged unit cells relax to slightly non-cubic structures, with the same relative energy difference therefore leading to an identical average SOC value compared to the cubic case. However the non-cubic phases are lower by ~>2 eV in both unit cell cases, explaining the large voltage differences between implementations and lattice geometries.

%(consult Table~\ref{tab:LFP-LTS},~\ref{tab:LCO},~\ref{tab:LNMO},~\ref{tab:Chevrel} in the SI for tabulated lattice parameters for all structures relevant for Figure \ref{fig:alignment})

%%% With all of these we get the alignment seen in Figure ...
\subsection{Final alignment across implementations}
\begin{figure}
    \centering
    \includegraphics[width=1\linewidth]{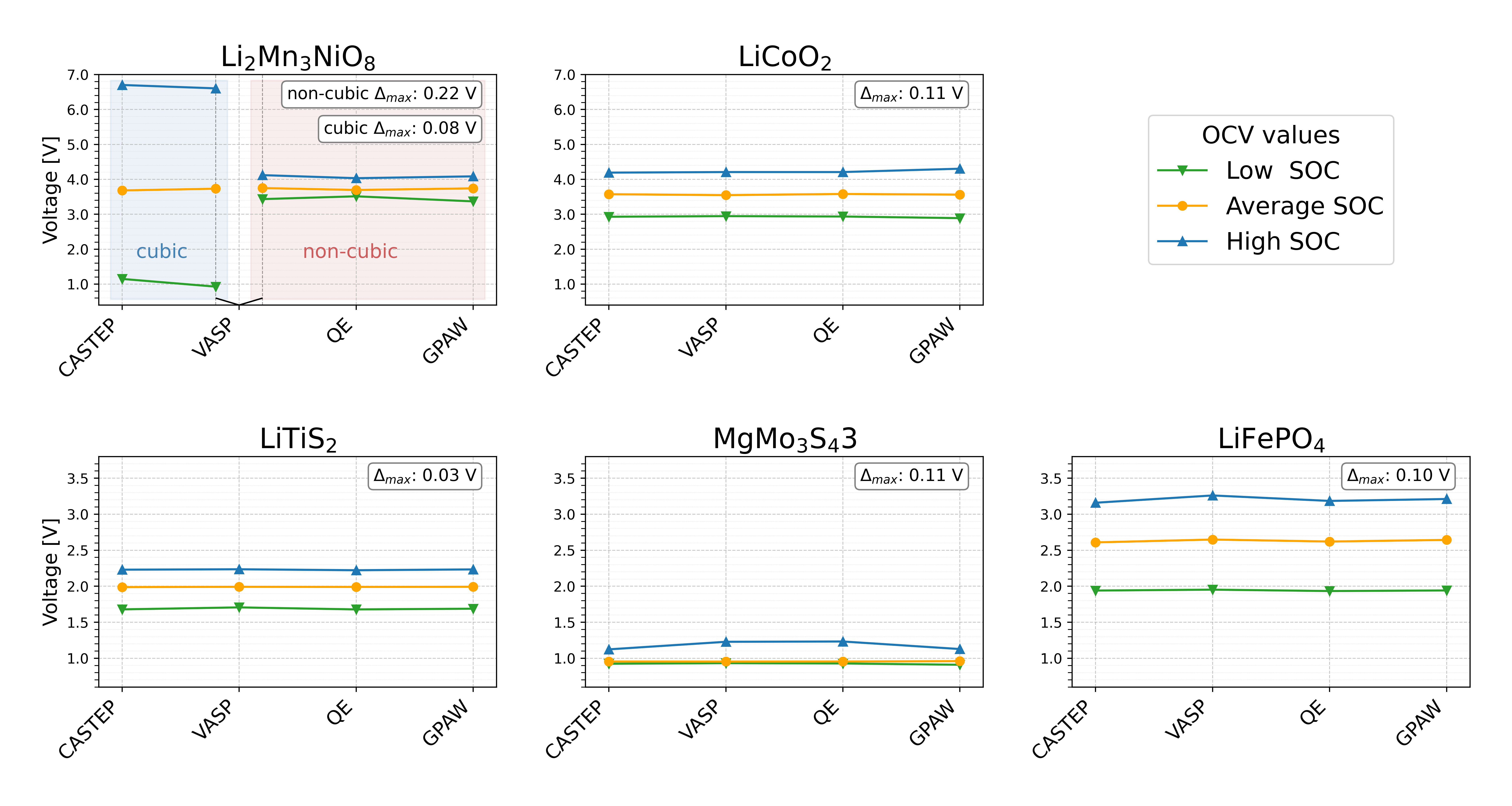}
    \caption{The OCV calculations for Li$_2$Mn$_3$NiO$_8$, LCoO$_2$, LiTiS$_2$, MgMo$_3$S$_4$, and LiFePO$_4$ across the DFT engines Quantum ESPRESSO, VASP, GPAW and CASTEP.}
    \label{fig:alignment}
\end{figure}

Figure~\ref{fig:alignment} summarizes the OCV results for the five implementations. When considering only the pristine charged and discharged unit cells, the average OCV values exhibit excellent alignment, with discrepancies of <$0.03$~V for LiTiS$_2$, Li$_2$Mn$_3$NiO$_8$, LiCoO$_2$, LiFePO$_4$, and MgMo$_3$S$_4$.

Achieving alignment beyond average OCV values becomes substantially more challenging when vacancies or single occupations within supercells are taken into account.
Across all Li-based systems, high SOC configurations generally exhibit better agreement across codes than low SOC configurations, as shown in Figure~\ref{fig:alignment}.

For instance, in the cubic Li$_2$Mn$_3$NiO$_8$ case, the lattice parameter of the cubic supercell computed with VASP and CASTEP differ by only $\approx 0.20\%$, and the OCV at high SOC is well aligned, implying that differences in the relaxed cell geometry are unlikely to be the primary cause of the misalignment at low SOC of about $\approx0.2$~V.
Instead, metastable states in low SOC structures appear to be the principal source of discrepancy, which is not apparent in the high SOC results. 
The near-perfect agreement for average OCVs derived from pristine cells further indicates that the main challenges are indeed associated with treating configurations involving fractional charge occupations and the resulting Fermi level shifts.
In particular, minimizing discrepancies for single charge-carrier occupations has proven easier than for single vacancy configurations. In the non-cubic case we similarly have the low and high SOC values exhibiting a larger spread, but in fairly good agreement.

As a sensitivity test, the well-aligned LiFePO$_4$ system was also run in VASP using the single-valence Li pseudopotential, instead of the three-valence Li\_pv pseudopotential used to achieve alignment. Interestingly, this yielded near-perfect alignment with the average- and high-SOC OCV values, but revealed a substantial discrepancy of approximately 0.4 V at low SOC. This underscores that, while some published pseudopotentials may perform well in many regimes, cases involving partial occupation can be particularly sensitive and lead to significant errors—highlighting the critical importance of pseudopotential choice.

It is important to emphasize that alignment across DFT codes is achievable but not trivial, particularly when implemented in an automated workflow. 
Procedures that are straightforward in focused material-specific studies, such as parameter tuning or manual inspection, become significantly more complex when scaled and automated, as is described in detail in Section \ref{sec:AlignmentChallenges}.
Inclusion of spin polarization, even when restricted to collinear orientations, introduces additional complications that can make otherwise routine steps problematic in an automated context \cite{ponet2024energy, xia2025electron}. 
Similarly, Hubbard corrections \cite{himmetoglu2014hubbard, qin2022hubbard} are often essential for an accurate description of redox activity in many battery cathode materials that involve strongly correlated 3\textit{d} electrons \cite{timrov2022accurate, hautier2011phosphates}.
Including these effects would further increase the complexity of alignment because they enlarge the energetic landscape to be optimized. For example by introducing further local minima in the potential energy landscape that arise due to the localization of electrons on different transition metal atoms. Furthermore, different DFT codes rely on different implementations with respect to projection schemes or cutoffs for magnetic moments and Hubbard corrections, posing further challenges.
For these reasons, the present study is limited to non-spin-polarized calculations at the GGA level without Hubbard corrections.
We note that addressing both spin polarization and Hubbard corrections consistently across codes is an essential direction for future work.

While the individual findings of this study are not entirely novel, it serves as a clear exemplification that misalignment across codes can arise from fundamentally different causes depending on the material, ranging from pseudopotential choices and convergence settings to optimizer algorithms and symmetry-breaking protocols, thereby greatly complicating the design and implementation of truly interoperable workflows.
The challenges and lessons learned in this work suggest design guidelines to consider for implementing DFT workflows and for aligning results across interoperable implementations. First, workflow design should explicitly incorporate lessons from validation. Cross-code comparisons are cumbersome, but they reliably expose misalignments and highlight errors that are otherwise hard to identify. In our case, lattice parameters were generally consistent, yet small structural differences could translate into large deviations in total energy derived properties. Second, differences in optimization algorithms should be considered to reduce the risk of trapping in local minima and to ensure robustness across implementations. Furthermore, standardized procedures across implementations for identifying local minima are important at apparent minima to mitigate symmetry trapping. Additionally, sharing final and/or intermediate structures between implementations allow for further relaxations that can help lead to an aligned minimum and total energetic properties.
Beyond benchmarking, cross-code validation should be viewed as a valuable control for complex calculations that are otherwise difficult to verify, helping to ensure reproducibility and scientific reliability.

\section{Conclusion} \label{conclusion}
This study addresses interoperability in automated DFT workflows by implementing the same OCV protocol across four workflow managers: AiiDA, PerQueue, Pipeline Pilot, and SimStack, which are coupled with CASTEP, GPAW, Quantum ESPRESSO, and VASP calculators. A general I/O structure for the OCV workflow was designed using JSON (with OPTIMADE-compliant structures), providing standardized definitions for inputs and outputs and enabling engine-agnostic execution and comparison. This standard design makes the scientific intent explicit, while allowing code-specific adapters to handle details of submission, relaxation, and parsing. 
Furthermore, the I/O facilitates FAIR practices by making results portable and machine-readable. The I/O JSON schema design also facilitates the easy integration of the workflow into an API-based MAP architecture, setting the stage for automated data exchange, streamlined interoperability between heterogeneous tools, and the possibility of scaling workflows across distributed research infrastructures.

This work demonstrates that a common JSON-based I/O with code-specific adapters enables cross-engine execution and comparison of an OCV protocol, clarifying where alignment succeeds and where it becomes challenging in practice. Across five implementations and multiple chemistries, voltages computed from pristine charged and discharged cells align closely - often within a few hundredths of a volt, while energetics from vacancy-containing supercells remain the principal source of disagreement.
Even under coordinated settings (PBEsol, non–spin polarization, shared k-point density, and stringent convergence criteria), final defect energetics are highly sensitive to workflow choices such as smearing scheme and width, symmetry handling, and relaxation procedures.
%Looking ahead, implementations should explicitly account for how tightly pristine cells are converged and propagate those tolerances to defect calculations. In this study, a seemingly minor change in a cubic lattice parameter of 0.0014 Å, resulting from a slightly smaller smearing width, coincided with a sharp drop in the vacancy structure's energy, underscoring that defect energetics can hinge on minute structural differences.
%and motivating the validation steps in automated workflows.

While OCV value calculations from different DFT codes can be brought into reasonable agreement across a range of cathode materials, achieving such alignment is far from straightforward, especially for workflows involving structures with vacancies.
Previous DFT benchmark studies have emphasized structural parameters of pristine materials; however, the OCV workflow inherently requires accurate total energies for both fully charged and discharged states, as well as for vacancy-containing supercells. To our knowledge, no benchmark study to date has systematically compared total energy-derived properties for vacancy structures across different DFT implementations; moreover, a clearly defined protocol for conducting such benchmarking studies has not yet been established. The study puts forward design guidelines for building robust, automated DFT workflows, with an emphasize on the value of cross-code implementations for validation to identify local minima, detect inadequate pseudopotential choices, and identify sensitive parameters, all to mitigate error propagation throughout the workflow.

% This studys show that seemingly similar convergence protocols can lead to distinct local minima and that cross-code comparisons provide a valuable means of validation of result. They also highlight the sensitivity of automated workflows to the optimization algorithm, starting structure and error propagation through the workflow.

Looking ahead, incorporating especially magnetism will be a necessary requirement to enable analysis of more complex materials and increasing the physical validity.
Although not explored here, future workflows should aim for including spin-polarization effects. This will add complexity in aligning +U conventions and magnetic moments across engines, but remains essential for realizing interoperability for realistic materials.

Beyond the JSON I/O, we publish a machine-readable description of the workflow as a JSON-LD graph of inputs, methods, results, and their provenance. Rather than ad‑hoc keys, we reuse controlled vocabularies and ontologies namely EMMO, BattINFO, and OSMO for scientific concepts, and Schema.org for resource descriptions. This ontological grounding clarifies field meanings across engines, enables machine-actionable queries, improves FAIRness, and promotes cross-domain interoperability.

\section{Code and data availability}
Dataset with I/O structures, semantic JSON-LD file and necessary DFT files for reproducibility:
\begin{itemize}
    \item https://doi.org/10.24435/materialscloud:e6-e4
\end{itemize}

Repositories for each implementation:
\begin{itemize}
    \item PerQueue/VASP and PerQueue/GPAW: https://gitlab.energy.dtu.dk/skste/ocv-perqueue-workflow-vasp-and-gpaw
    \item PipelinePilot/CASTEP: Proprietary
    \item AiiDA/Quantum ESPRESSO: https://github.com/tsthakur/aiida-open\_circuit\_voltage
    \item SimStack/VASP: All code and documentation related to this OCV workflow using the SimStack framework are publicly accessible in our GitHub https://github.com/KIT-Workflows/Battery-Electrodes.
\end{itemize}

%%% Repos of the implementations, collect the i/o jsons in an entry to a data repo.

\section{Acknowledgements}
This project received funding from the European Union’s Horizon 2020 research and innovation program under grant agreement no. 957189 (BIG-MAP) and no. 101103873 (UltraBat).
The authors acknowledge BATTERY2030PLUS, funded by the European Union’s Horizon 2020 research and innovation program under grant agreement no. 957213.
S.K.S., M.D., J.M.G.L, T.V. and I.E.C acknowledge support from the Novo Nordisk Foundation Data Science Research Infrastructure 2022 Grant: A high-performance computing infrastructure for data-driven research on sustainable energy materials, Grant no. NNF22OC0078009.
T.S.T., N.M. and G.P. acknowledge support from the NCCR MARVEL, a National Centre of Competence in Research, funded by the Swiss National Science Foundation (grant number 205602), along with support from the Swiss National Supercomputing Centre (CSCS) under project ID mr32 (EIGER). G.P. acknowledges support by the Open Research Data Program of the ETH Board (project ``PREMISE'': Open and Reproducible Materials Science Research). T.V. acknowledge support for the Pioneer Center for Accelerating P2X Materials Discovery (CAPeX), DNRF grant P3. 
W. W. and C. R. C. R. thank the German Federal Ministry of Education and Research (BMBF) for financial support of the project Innovation-Platform MaterialDigital (\url{www.materialdigital.de}) through project funding FKZ number 13XP5094A.

%%%%% EVERYBODY %%%%% add acknowledgements

\printbibliography

\begin{refsection}

\subfile{SupplementaryInformation}

\end{refsection}

\end{document}

%% file: SupplementaryInformation.tex
\title{Supporting Information - Challenges in Creating and Aligning Interoperable DFT Workflows Across Implementations}

\maketitle

\begin{center}
  \vspace{-2.5em} % adjust vertical spacing if you like
  \normalsize\textbf{$\dagger$These authors contributed equally to this work:}\\[0.3em]
  \normalsize\textbf{Simon K. Steensen and Tushar Singh Thakur}
  \vspace{1.0em}
\end{center}

%\newpage
%\setcounter{section}{0}
%\setcounter{figure}{0}
%\setcounter{table}{0}
%\setcounter{equation}{0}

\setcounter{section}{0}
\renewcommand{\thesection}{S\arabic{section}}
\makeatletter
\renewcommand*{\theHsection}{SI.\arabic{section}}
\makeatother

\renewcommand{\thesection}{S\arabic{section}}
\renewcommand{\thefigure}{S\arabic{figure}}
\renewcommand{\thetable}{S\arabic{table}}
\renewcommand{\theequation}{S\arabic{equation}}

\section{I/O structure}\label{SI:sec:IO}
\subsection{Common Interface across implementations}
To ensure interoperability among different workflow managers and DFT engines, a common input and output (I/O) interface has been developed. 
This interface adopts a structured JSON/YAML based schema that standardises both the format of input files taken by various DFT engines and the format of the outputs they produce.
By enforcing a consistent data model, all implementations are constrained to adhere to the same conventions, ensuring reproducibility and ease of integration within Materials Acceleration Platform (MAP) frameworks.

\paragraph{Input format}
The input file follows a top-level JSON dictionary of the form: 
\{\texttt{task}: \texttt{ocv}, \texttt{inputs}: INPUTS, \texttt{meta}: META\}. 
where \texttt{task} specifies the type of workflow to be executed, with the current work only supporting \texttt{ocv} (open-circuit voltage) workflow, \texttt{inputs} is a dictionary containing the parameters required for the calculation as defined below, and \texttt{meta} is an optional dictionary for engine-specific metadata, such as the choice of quantum engine or computational resource allocation.
If \texttt{meta} is omitted or empty, the workflow will assign default values and produce a valid, reproducible result. 

The following inputs are required for a valid workflow execution:
\begin{itemize}
    \item \texttt{structure}: A structure provided in JSON format following the OPTIMADE specification \cite{andersen2021optimade}. The dictionary must contain the keys: \texttt{source}, \texttt{species}, \texttt{\seqsplit{dimension\_types}}, \texttt{lattice\_vectors}, \texttt{cartesian\_site\_positions}, and \texttt{species\_at\_sites}, corresponding to the content typically found under \texttt{["data"]["attributes"]} in OPTIMADE compliant data structures. Remaining OPTIMADE keywords may be provided as per the discretion of the user, but are not required.
    \item \texttt{protocol}: A string specifying the computational accuracy level, one of \texttt{fast}, \texttt{moderate}, or \texttt{precise}. Each level corresponds to an internally mapped set of DFT parameters controlling thresholds for total energy and force convergence, k-point sampling, self-consistency tolerances, volume relaxation criterion, supercell construction, and pseudopotential selection. The \texttt{precise} protocol is recommended for cross-engine comparison.
    \item \texttt{engine}: A string identifying the simulation engine and optionally the computational resource. Valid options are retrieved from a \texttt{get\_valid\_engines()} function implemented by each workflow system. This function returns a JSON dictionary where each key corresponds to a valid engine identifier, and its value is a nested dictionary with at least a \texttt{description} field. Example: {“qe-7-daint”: {“description”: “Quantum ESPRESSO pw.x v7.0, GPU accelerated, running on the CSCS Piz Daint machine”}, “abinit-eiger”: {“description”: “ABINIT on CSCS Eiger machine, using PseudoDojo pseudos v0.4”}}
\end{itemize}
Optional parameters to provide finer control over the simulation setup, allowing users to override protocol defaults.
Following is an exhaustive list of optional inputs allowed:
\begin{itemize}
    \item \texttt{magnetization\_treatment}: A string which can be \texttt{none}, \texttt{collinear}, or \texttt{\seqsplit{noncollinear}} (default: \texttt{none}), specifying how spin polarisation is treated. Magnetisation may evolve during the workflow, for instance in NEB calculations, so we recommend extra caution in such cases.
    \item \texttt{spin\_orbit}: Boolean flag enabling spin–orbit coupling (default: \texttt{false}).
    \item \texttt{magnetization\_per\_site}: List of floats in Bohr magnetons, defining the initial magnetic moments on each atomic site. Must have the same length as the number of atomic sites (default: a list of zeros).
    \item \texttt{kpoints\_mesh}:  Explicit k-point grid, specified as: 
        \texttt{\seqsplit{"mesh": [nx, ny, nz], "offset": [ox, oy, oz]}}
    where \texttt{nx, ny, nz} are integers defining the Monkhorst–Pack grid size and \texttt{ox, oy, oz} are fractional offsets (commonly \texttt{[0.5, 0.5, 0.5]}). Either this or \texttt{kpoints\_distance} must be provided.
    \item \texttt{kpoints\_distance}: Minimum reciprocal-space distance (in \AA$^{-1}$) between k-points (default: 0.15 \AA$^{-1}$). Used to generate the mesh automatically if \texttt{kpoints\_mesh} is not given.
    \item \texttt{volume\_change\_stability\_threshold}: Relative volume change threshold (default: 10 \%). If the volume difference between charged and discharged structures exceeds this limit, the workflow is terminated, flagging the system as structurally unstable.
    \item \texttt{cation}: Chemical symbol of the charge carrier (default: "Li").
    \item \texttt{supercell\_distance}: Minimum image–image separation distance (in \AA) used for supercell construction (default: 8 \AA).
    \item \texttt{bulk\_cation\_structure}: The bulk structure of the cation for reference energy computation, provided in the same JSON format as structure (default: internally stored bulk Li structure).
\end{itemize}

\paragraph{Output format}
The output format follows a structure similar to the input, with a top-level JSON dictionary of the form:
{\texttt{task} \texttt{ocv}, \texttt{outputs}: OUTPUTS, \texttt{meta}: META}.
where \texttt{task} identifies the type of workflow that has been executed (currently, only \texttt{ocv} is supported), \texttt{outputs} is a dictionary containing all results whose common format is specified below, and \texttt{meta} serves as an optional container for engine-specific metadata (for example, the code or workflow version used). 
Although \texttt{meta} may be empty, it provides a flexible mechanism for storing auxiliary information relevant to a particular implementation. 
It is mandatory that all quantities defined in the common specification be reported under the \texttt{outputs} dictionary, even if they are also included in the meta section.

The \texttt{outputs} dictionary must contain the key \texttt{OCV\_values\_V}, a nested dictionary holding the primary results of the OCV workflow described as follows:

\begin{itemize}
    \item \texttt{OCV\_average}: The computed average OCV value, in Volts.
    \item \texttt{OCV\_low\_SOC}: The OCV value at the low state of charge, i.e., when the first cation is removed. This field is optional and must only be included if the volume change remains below the threshold defined by \texttt{volume\_change\_stability\_threshold}.
    \item \texttt{OCV\_high\_SOC}: The OCV value at the high state of charge, i.e., when the last cation is removed. This field follows the same conditional inclusion rule as above.
    \item \texttt{fully\_discharged\_structure}: The fully relaxed discharged structure, represented in JSON format and compliant with the OPTIMADE specification.
    \item \texttt{fully\_charged\_structure}: The fully relaxed charged structure, in the same OPTIMADE compliant format.
    \item \texttt{fully\_charged\_structure\_with\_discharged\_cell}: The relaxed charged structure constrained to the discharged cell geometry, obtained by scaling atomic positions and lattice vectors according to the ratio of the charged to discharged volumes.
    \item \texttt{low\_SOC\_structures} and \texttt{high\_SOC\_structures}: Lists of relaxed supercell structures corresponding to the low and high SOC configurations, respectively. These are optional and should only be reported when the volume change criterion is satisfied.
\end{itemize}

The following quantities provide additional information derived from the DFT calculations and are reported if and when available:

\begin{itemize}
    \item \texttt{total\_energies}: A dictionary containing the total energies (in eV) for each structure, with keys \texttt{fully\_discharged}, \texttt{fully\_charged}, \texttt{low\_SOC}, and \texttt{high\_SOC}. The absolute energy reference may vary across implementations, but energy differences must be consistent and physically meaningful.
    \item \texttt{forces}: A dictionary providing the atomic forces (in eV/\AA) for each structure, with the same keys as \texttt{total\_energies}. Each entry contains an N×3 list of Cartesian force components.
    \item \texttt{stresses}: A dictionary of stress tensors (in eV/\AA$^{3}$) for each structure, each represented as a 3×3 list of Cartesian components.
    \item \texttt{total\_magnetizations}: The total magnetisation (in $\mu_\mathrm{B}$) for each structure. This field is optional but must be included when the input parameter \texttt{\seqsplit{magnetization\_treatment}} is not set to \texttt{none}.
\end{itemize}

Implementations may include auxiliary quantities in the meta section for transparency or debugging purposes. 
Additionally, a utility function is provided to compute the percentage volume change between \texttt{fully\_charged\_structure} and \texttt{\seqsplit{fully\_charged\_structure\_with\_discharged\_cell}}. 
This serves as an internal consistency check and determines whether subsequent supercell calculations should proceed, as defined by the \texttt{volume\_change\_stability\_threshold parameter} in the input schema.

\section{Semantic description}\label{SI:sec:sematics}
\subsection{Machine-readable context}

The semantic description of a resource (e.g., data, metadata, input files) involves linking the parts of the resource to a controlled vocabulary. In this way, both humans and computers can understand the meaning attributed to the resource and its parts, in an unambiguous way.

We use terms available in authoritative, EU-wide vocabularies to describe physical, chemical and battery concepts:
\begin{itemize}
    \item The Elementary Multiperspective Materials Ontology (EMMO)\cite{EMMO}, a foundational ontology of concepts describing the physical world: particles, space-time, materials and observables.
    \item The Battery Interface Ontology (BattINFO)\cite{BattINFO, clark2025semantic, clark2022toward}, a set of domain ontologies describing concepts on batteries and electrochemistry.
    \item The Ontology for Simulation, Modeling, and Optimization (OSMO)\cite{OSMO}, a domain ontology associated to the the MOdelling DAta (MODA) standard that defines terms and templates to document the simulation of materials.\cite{MODA}
\end{itemize}

Each term is uniquely identified with an Internationalized Resource Identifier (IRI) \cite{IRI}, a sequence of characters within the UTF-8 character encoding that uniquely describes a web resource. Concepts are also annotated with human-readable definitions, labels, and notes to make their meaning explicit for humans. 

Descriptions of physical and chemical domains are complemented with additional vocabularies and ontologies:
\begin{itemize}
    \item The XML Schema Definition (XSD) \cite{W3}, the recommendation from the W3C consortium on how to describe elements in an extensible markup language (XML) document, including data types.
    \item The Web Ontology Language \cite{OWL} with concepts to represent knowledge into classification networks. 
    \item The Resource Description Framework \cite{RDF}, which provides a set of syntax notation and formats to describe graph data in a machine-readable format.
    \item The Schema.org vocabulary, \cite{schema-org} with terms that describe web resources (locations, events, multimedia, agents, etc.). 
\end{itemize}

\subsection{General Procedure}

Although there are multiple ways one could describe a workflow, it is best practice to re-use authoritative resources and guidelines. The MODA template already enforces a template for the description of simulations; therefore, it is natural to use the MODA template and its semantic description in OSMO to build the semantic description of the workflow. We follow the general steps:

\begin{itemize}
    \item The domain expert fills the MODA template as comprehensively as possible.
    \item With the template, the ontologists then select the ontology concepts that best describe the keys and values of the template.
    \item After this selection, the ontologists encode the relationships between the template and the concepts in a semantic format.
    \item The onotologist and domain expert discuss the proposal to ensure that the descriptions and links reflect the meaning given to the resource by the domain expert.
\end{itemize}

\subsection{Implementation}

We have selected the \cite{JSONLD} format, as it is open source, intelligible to humans (semi-structured ASCII text), similar to JSON, widely used to exchange data, and designed to link data. Like JSON, a JSON-LD document stores key-value pairs in text format forming hierarchies, but in addition enforces additional syntax designed to support the description of linked data:

\begin{itemize}
    \item The document must start with a  \texttt{"@context":{...}} object, specifying the namespaces of the vocabularies that will be used.
    \item Each object must have a \texttt{"@type":"<IRI>"} pair, specifying the concept the object maps to. 
    \item Each object must have a \texttt{"@id":"<URI>"}  pair, assigning a unique identifier to the object. If none is provided, a default is created.
    \item Objects nest into a hierarchy of semantic relationships.
\end{itemize}

\begin{figure}[h!]
    \centering
    \includegraphics[width=1.0\textwidth]{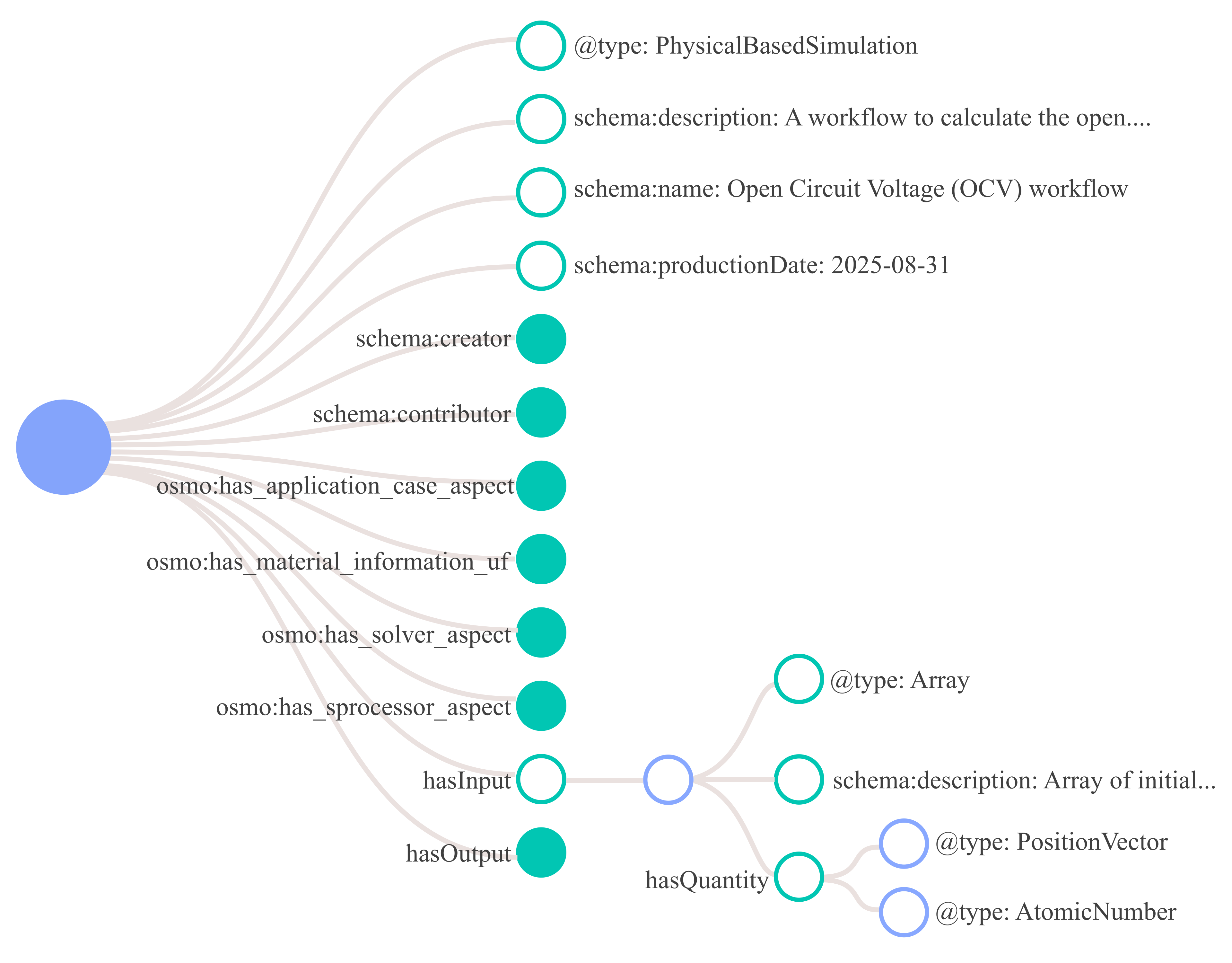}
    \caption{Hierarchical representation of the JSON-LD description of the workflow. Purple circles represent objects (nodes), green open circles are attributes (edge-node), while green filled circles nest more objects that are not shown for clarity. Attributes with the prefix \texttt{schema:} and \texttt{osmo:} are described in the Schema.org vocabulary and OSMO ontology, respectively. Attributes without prefix inherit from the EMMO ontology. The uncollapsed \texttt{hasInput} links to an object of type \texttt{Array} with a human-readable description \texttt{schema:description} and the quantities within the array \texttt{PositionVector} and \texttt{AtomicNumber}. }
    \label{fig:jsonld}
\end{figure}

The workflow is described as a JSON-LD object, with the same parts as in the MODA template. These parts branch into methods, procedures, measurements and quantities, each described with an appropriate vocabulary term or ontology concept. A representation of the workflow description is illustrated in Figure \ref{fig:jsonld}. The current version of the JSON-LD description is included in the Supplementary Materials.

\section{DFT data}
This section compiles the data underpinning the DFT relaxations and OCV calculations. We first list the code-specific DFT settings used across engines (CASTEP, GPAW, Quantum ESPRESSO, VASP), including.
Following this are the hard-coded scalings used for the supercells for each material.
We then report the relaxed lattice parameters for all structures employed in the final OCV alignment (charged, discharged, charged-constrained, and the low/high SOC supercells).

Next, we present the smearing convergence tests carried out for LiFePO$_4$. Finally, we provide projected densities of states (PDOS) for the investigated low-/high-SOC configurations of LiFePO$_4$ and Li$_2$Mn$_3$NiO$_8$, used to interpret residual discrepancies across codes. Code and data availability are linked in the main text.

\subsection{Parameters used for the DFT engines}
Across all implementations the PBEsol functional was used with a k-points density of 0.15~\AA$^{-1}$. The smearing implementations had to vary across due to inaccuracy of results. All the DFT code specific files can be found at https://doi.org/10.24435/materialscloud:e6-e4.

\subsubsection{VASP}
The high fidelity parameters used in the two VASP based implementations using SimStack and PerQueue as queueing tools were aligned to be identical. The parameters set were:
encut: 520.0 eV, ediff = 1.00e-07 eV, ediffg: -1.00e-03 eV, Symprec: 1.00e-08, algo: normal, prec: accurate, ibrion: 2 (ibrion: 1 if the calculation is stalled and close to the ground-state), isym: 2, nelm: 350, nelmin: 8, nsw: 250.

\subsection{Quantum ESPRESSO}
The pseudopotentials were taken from the standard solid-state pseudopotential libraries (SSSP) ~\cite{natureQE} for high-precision materials modelling (SSSP precision). The plane-wave cutoff energy was set to ecutwfc: 90 Ry, and ecutrho: 1080 Ry for the charge density cutoff. For the electronic convergence threshold 1.00e-08 Ry was chosen. For ionic relaxation, the force convergence threshold was set to 1.00e-04 Ry/Bohr and total energy convergence was set to 1.00e-05 Ry.

\subsubsection{GPAW}
The GPAW parameters used in the PerQueue/GPAW implementation set using ASE calculators were:
mode: pw-ecut 1200 eV, eigensolver: dav-niter of 6, maxiter: 500, energy-convergence: 0.0005 eV, density-convergence: 0.0001 e, eigenstates-convergence: 4e-08 eV, bands: occupied, 'forces': 0.001 eV/Å. The standard pseudopotentials generated with the \texttt{gpaw setup} command are used for all materials.

\subsubsection{CASTEP}
The standard, on-the-fly generated ultrasoft pseudo potentials were used in all CASTEP calculations. The protocol settings in the json-file were mapped to the \emph{Quality Settings} used in Materials Studio Collection for Pipeline Pilot. \texttt{Fast} was mapped to \emph{Coarse}, \texttt{moderate} to \emph{Medium} and \texttt{precise} to \emph{Ultra-Fine}. The production runs with CASTEP were executed using the \texttt{precise} protocol. The \texttt{precise} settings for CASTEP were: the energy cutoff for the planewave basis set was 630 eV, k-point separation of 0.04 ~\AA$^{-1}$ and the convergence criteria for the SCF cycle was 10$^{-9}$ eV. The convergence criteria for the Geometry Optimization were maximum energy difference: 10$^{-6}$ eV, maximum force difference: 0.01 eV/~\AA, maximum stress difference: 0.01 eV/~\AA.

\subsection{Scaling for in each material}\label{SI:scaling}
We define fixed scaling matrices for all materials and implementations to ensure direct cross-code comparability via identical supercell constructions. Establishing a universally applicable protocol for supercell generation remains non-trivial, as it must be compatible with multiple engines, coding languages and their symmetry/convention differences. Moreover, small deviations in relaxed unit cells across codes can propagate into divergent supercell shapes and sizes. We therefore specify the scaling matrices explicitly here to enforce consistent supercell sizes across implementations for testing the five materials at hand.
The matrices for all materials are:

\[
\text{Li$_2$Mn$_3$NiO$_8$ : }
\begin{bmatrix}
1 & 0 & 0 \\
0 & 1 & 0 \\
0 & 0 & 1
\end{bmatrix}   \qquad \text{MgMo$_3$S$_4$}: \begin{bmatrix}
2 & 0 & 0 \\
0 & 2 & 0 \\
0 & 0 & 2
\end{bmatrix} \qquad 
\text{LiTiS$_2$: }
\begin{bmatrix}
3 & 0 & 0 \\
0 & 0 & 2 \\
0 & 3 & 0
\end{bmatrix}
\]\\

\[
\text{LCoO$_2$: }
\begin{bmatrix}
1 & 0 & 0 \\
0 & 1 & 0 \\
0 & 0 & 1
\end{bmatrix}   \qquad \qquad \text{LiFePO$_4$}: \begin{bmatrix}
2 & 0 & 0 \\
0 & 0 & 1 \\
0 & 3 & 0
\end{bmatrix}
\]\\

\subsection{PDOS of Li$_2$Mn$_3$NiO$_8$ low soc in VASP at different smearings}
\begin{figure}[htbp]
    \centering
    \begin{subfigure}[b]{\textwidth}
        \centering
        \includegraphics[width=0.8\textwidth]{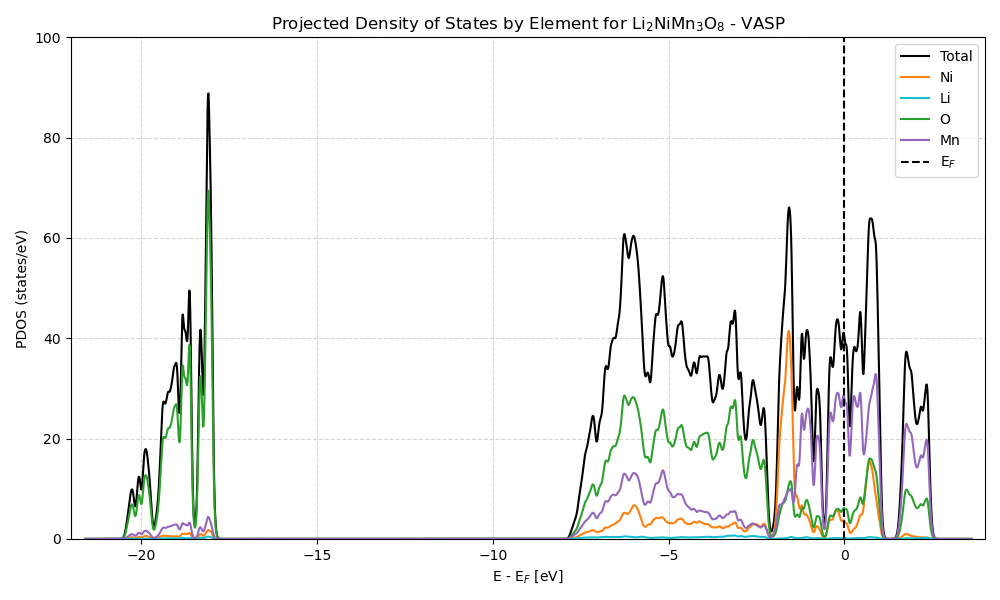} % Replace with your image file
        \caption{The projected density of states for the low SOC structure at 0.029 eV Fermi-Dirac smearing width produced by VASP}
    \end{subfigure}
    
    \vspace{1cm} % Adjust spacing between the figures
    
    \begin{subfigure}[b]{\textwidth}
        \centering
        \includegraphics[width=0.8\textwidth]{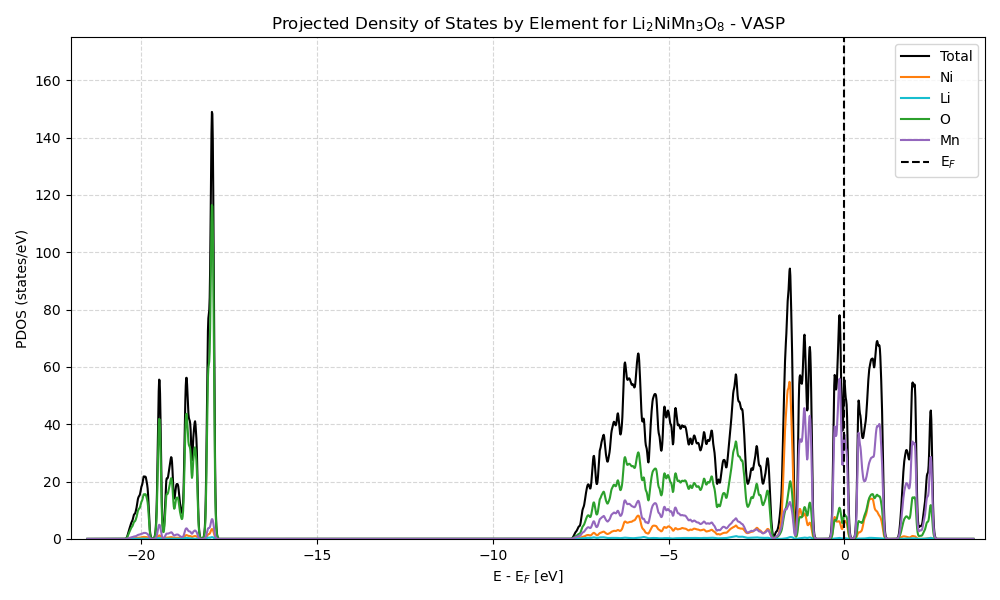} % Replace with your image file
        \caption{The projected density of states for the low SOC structure at 0.019 eV Fermi-Dirac smearing width produced by VASP}
    \end{subfigure}
    
    \caption{The two PDOS for the low SOC structure in the workflow run with Li$_2$Mn$_3$NiO$_8$. The PDOS reveal very different electronic structures for the two cases}
    \label{fig:PDOS_LNMO}
\end{figure}

To diagnose the non-monotonic low-SOC OCV behavior observed for cubic Li$_2$Mn$_3$NiO$_8$, we computed the PDOS for the VASP low-SOC structure at the two smearing widths at the abrupt drop in the low-SOC voltage. As discussed in the main text, the PDOS in Figure ~\ref{fig:PDOS_LNMO} shows a smearing-induced redistribution of fractional occupations near the Fermi level that produces an apparent gap to the next manifold of states at intermediate smearing, correlating with the energy drop. When using the tetrahedron method with Blöchl corrections (ISMEAR = -5) this artificial gap closes, supporting the interpretation that the observed discontinuity stems from the occupation scheme rather than a genuine change in electronic structure.

\newpage
\printbibliography